\documentclass[aps,prl,twocolumn,showpacs,superscriptaddress,groupedaddress]{revtex4}  
\usepackage{graphicx}  
\usepackage{dcolumn}   
\usepackage{bm}        
\usepackage{amssymb}   

\begin{document}

\bibliographystyle{unsrt}


\title{Effect of light polarization on plasma distribution and filament formation }

\author{L.Arissian$^{1,2}$ ,D.Mirell$^{2,3}$, J.Yeak$^{2,3}$, S. Rostami$^{2,3}$ and J.-C. Diels$^{1,2,3,}$ \\
$^1$ Electrical and Computer Engineering, University of New Mexico, USA \\
$^2$  Physics and Astronomy, University of New Mexico, Albuquerque, New Mexico, USA \\
$^3$ Center for High Technology Materials, 1313 Goddard SE, Albuquerque, New Mexico 87106}
\email{ladan@unm.edu}
\date{\today}

\begin{abstract}
We show that, for 200 fs light pulses at 790 nm,  the formation of filaments is strongly affected by the laser light polarization .
 Filamentation does not exist for a pure circularly polarized light, propagating in vacuum before focusing in air,
 while there is no difference for focusing the light in air or vacuum for linearly polarized light.
\end{abstract}


\maketitle


For a laser beam of sufficient intensity, air itself is a nonlinear medium, having a complex intensity
dependent index of refraction, nonlinear absorption, induced birefringence,
and becoming a partially conductive medium. These properties
lead to light filamentation, a situation where the nonlinear properties of
air determine the propagation properties.  While they have been investigated
since their discovery in 1995 both in the IR~\cite{Braun95} and the UV~\cite{Zhao95},
their formation and nature are still the object of controversy~\cite{Dubietis04,Diels10}.

In strong field ionization the light polarization had been extensively investigated~\cite{Faisal72}
and used to control re-collision of the electron~\cite{Dietrich94,Arissian10}.
Recently~\cite{Zhou11} detailed measurements of current in argon and nitrogen at filament intensities
showed a strong dependence on light polarization. Measurement presented here demonstrate
the effect of light polarization in filament formation.
This effect has not been clearly investigated previously, because of the difficulty in
1) ascertaining that the beam has the expected polarization at the point where filamentation starts~\cite{Close66}
and 2) making polarization measurements of a high intensity filament.
These difficulties relate to the fact that the filamentation study is not a chamber study
like most of the strong field light-matter interactions.
Although there had been evidence of polarization deformation of high intensity beam propagating  in air~\cite{Close66},
filaments are prepared over a long propagation distance in gases.
In order to have a clear starting point and a well defined initial condition for the filament,
 we use an aerodynamic window to prepare the filament in vacuum before launching it in air.
 By focusing the beam in vacuum to the filament size, all pre-filament nonlinear effects are eliminated~\cite{Arissian12c}.
 Given such initial condition no filament is generated with circularly polarized light prepared in vacuum,
while filaments are observed with the same circularly polarized beam propagating in air in the pre-filamentation stage.
 For the linearly polarized light filaments persist in both cases of pre-propagation in air and vacuum, with filaments
 prepared in vacuum  having 20 percent longer length.

An infrared filament is an ideal object to study strong
field light-matter interaction, in which light and matter have a mutual
recordable effect on each other, resulting in a confinement of a laser beam
in a few hundred micron diameter channel, over distance, in excess of the Rayleigh range
for that beam size.

 For the few hundred fs long pulses intense enough to create a filament in air,
 the interaction of light is with bond electrons in
atoms or molecules, free electrons created by tunnel ionized, and partially
orientated molecules. Since the modification of light happens in a time
scale much faster than a plasma period, a careful microscopic (in the fs
scale) study of the parameters involved in filament formation is needed.

There has been a
considerable number of models and numerical simulations on filaments.  While each of these simulations
attempts to explain a particular observation, a physical picture fails to emerge. The existing formalism
lacks for instance details on the effects of plasma on the wavefront and light polarization at the microscopic level.
In the experiment also, there is not a direct method to image (single shot) the light propagation, or control pre-filament
light environment. 
The controversy in filament studies can be monitored in the optics literature with complete opposite statements such as
{\em ``a filament results from balance between Kerr self-focusing and plasma defocusing''}~\cite{Couairon07} or
{\em ``a filament is the manifestation of a moving focus''}~\cite{Shen71,Brodeur97}, or  {\em ``a filament is a
Bessel beam''}\cite{Dubietis04}, to the extent that there is no more a clear definition for filamentation.
Even clarification between optical breakdown and filamentation seems to be necessary~\cite{Nguyen03}.
One reason is the shear number of physical phenomena associated with the filaments, such
as conical emission, THz emission, harmonic generation, spatial replenishment~\cite{Mlejnek98}
self-healing~\cite{Kolesik04}, {\em etc}.
Another reason is that there is an obscure
curtain of the order of several meters that separates the prepared initial condition from the observed filament.
Substantial temporal~\cite{Bernstein02} and spatial~\cite{Grow06} reshaping of the initial pulse profile takes place in amplitude and
phase, such that the pulse  that ultimately reaches a self-focus of the size of the filament
is very different from the macroscopic beam launched in the atmosphere.

Filaments have traditionally been created by letting a high power laser self-focus in air.
Besides losing a lot of energy during that process, the starting point of the filament is undefined.
Furthermore, it becomes difficult, sometimes impossible, to distinguish the phenomena associated with the ``preparation phase''
from the true ``filamentation phase''.  We focus instead the beam in a vacuum cell with a 3 m focal distance lens,
to a  beam waist at the location close to the transition of vacuum to air.  The intensity at the focal spot
in vacuum exceeds 1 TW/cm$^2$.  At those intensities any window material is either damaged or shows significant nonlinear effects.
 The aerodynamic window provides a pressure gradient across a supersonic air stream, with an
 expansion chamber profiled in such a way that the pressure on one side is atmospheric,
 and on the other side less than 10 torr. The contours of the window are such that a
 pressure gradient is formed in the supersonic flow by Prandtl-Meyer~\cite{Liepmann57}
 expansion waves across which the beam propagates out of the vacuum chamber.
 The supersonic gas flow enters the diffuser, recovers the flow pressure back to atmospheric conditions, and ejects into the atmosphere.
 The supply pressure upstream of the supersonic nozzle can be varied to achieve optimum performance.
 The compressor associated with this window supplies a continuous flow of 10 m$^3$/min (150 dm$^3$/s) at a pressure of 8 kg/cm$^2$.

We tackle for the first time one of the major obstacles of the filamentation studies; controlling the pre-filamentation
 light propagation.  By focusing the beam in vacuum  to the filament size before launching in the atmosphere (through the aerodynamic window),
 all pre-filament nonlinear effects are eliminated.  With such a  propagation in vacuum up to the focus,
 filaments are observed only with linear polarization.  No filament is observed with circularly polarized light,
 in apparent contradiction with the theoretical simulation of  Panov et al~\cite{Panov11}
 where it is claimed that filaments would be more uniform and intense with circularly polarized light.

To study the effect of polarization on filament formation, the output of a home-build amplified Ti:Sapphire
laser with 10 mJ pulse energy and 200 fs pulse duration is focused with a 3m focal length lens. The conductivity
due to the generated free electrons and the fluorescence of nitrogen are measured as a function of distance
from the geometrical  focus for linearly and circularly polarized light prepared either in vacuum or air.
In linear polarization, measurements of the beam profile versus distance from the aerodynamic window have
indicated that filaments are produced  in both vacuum and air pre-filament cases.
These measurements employed for the first time a {\em linear} attenuator consisting of a
coated glass at 89$^{\rm o}$ incidence (attenuation factor $> 10^8$)~\cite{Diels10}.
In all cases where a filament is produced, its radius ($1/e^2$ of the intensity, or FWHM/1.177) is measured to be
w$_0$ = 250 $\mu$m.  It should be noted that, in the case of preparation in
vacuum, the 1/e$^2$ half-width at the focus (in vacuum) is w$_0$ = 150$\mu$m.
With no attenuation before reaching the air boundary, the intensity
launched in the atmosphere in the case of the vacuum preparation is  $2.8 \times 10^{14}$ W/cm$^2$,
considerably higher than the ``clamped intensity'' of $3 \times 10^{13}$ W/cm$^2$ in a filament~\cite{Daigle10}.

\begin{figure}[h!]
\includegraphics[scale=0.33]{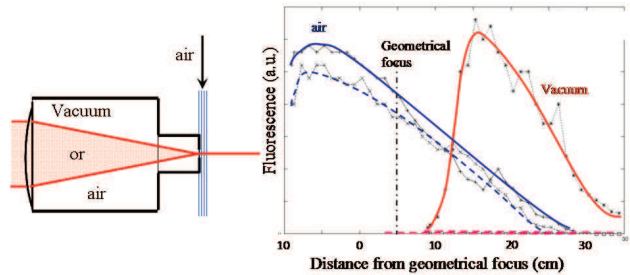}
\caption[]{\small Left: a 15 mm diameter beam is focused in a 3 meter long vacuum tube,terminated by an aerodynamic window.
Right: The fluorescence at 337 nm is measured as a function of distance from the geometrical focus,
for linearly polarized light (solid lines) and circularly polarized light (dashed lines).}
\label{fig:fluorescence}
\end{figure}

\begin{figure}[h!]
\includegraphics[scale=0.35]{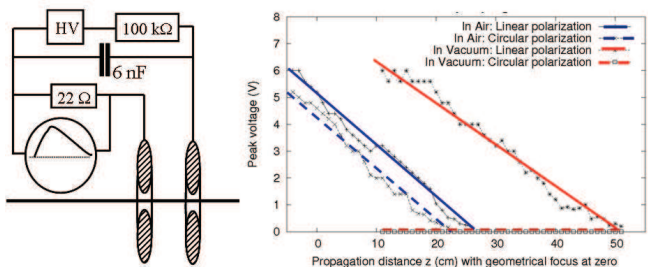}
\caption[]{\small Left: circuit to measure the conductivity. The filament is sent
through two profiled holes in electrodes 2 cm apart. Air becomes conductive under passage
of the filament, and a voltage is measured across a 22 $\Omega$m resistor.
Right: measured voltage (hence conductivity) as a function of distance
from the geometrical focus of a 3 m focal distance lens.
The solid lines correspond to linear polarization; the dashed lines to circular polarization.}
\label{fig:conductivity}
\end{figure}

A plasma is associated to these filaments, which manifests itself through fluorescence of positive ions,
and an increased conductivity of air.
Our measurement traces the fluorescence of nitrogen at 337 nm (Fig.~ \ref{fig:fluorescence}),
as a function of distance from the geometrical focus.
In the case of linearly polarized light focused in vacuum, the fluorescence rises
and decays over approximately 30 cm.  Measurements of the beam
profile versus distance however shows the filament to conserve a
constant diameter for more than 75 cm~\cite{Kolesik10,Diels10}.
When the beam is filamented in air, the plot
of fluorescence versus distance is receded by  about 20 cm, which
represents the difference between  the linear and nonlinear focus.
There is very little difference in fluorescence for the initially
circularly polarized beam self-focusing in air (dashed line), as compared to the case of  linear polarization. In contrast,
there is no fluorescence observed in the case of circularly polarized beam
started from vacuum.
The absence of plasma in circular polarization is confirmed by conductivity
measurements presented in Fig.~\ref{fig:conductivity}, where the conductivity of plasma
induced by the filaments is plotted as a function of distance from the geometric focus.

The explanation of our observation lies in the microscopic nonlinear absorption of electrons.
For sub picosecond laser filamentation, the generated plasma is far from equilibrium
and the time scale of the interaction --- modification of light through interaction with ionized medium ---
is much shorter than a plasma period.
In classical models of fluctuations of plasma density~\cite{Jackson75}
there is no distinction between the response of electrons generated by circularly polarized light
and those generated by linearly polarized light.
The difference  in ionization rate between circular if
and linear polarization~\cite{Krainov97, Ammosov97} is not sufficient to account for the
absence of filament when preparing circularly polarized filaments in vacuum.
When prepared in air, we observed filamentation with circular and linear polarized light
at the same pulse energy.

We chose to consider filamentation as a strong field ionization process where
light and matter have equal share. Adopting the tunnel ionization model the tunneled electrons
are not isotropic and the direction of electron ejection is set by light polarization at the moment of
ionization, while the trajectory of the motion is determined by the laser field~\cite{Corkum93}.
In this approach the plasma dynamic is set by laser light rather than the density of electrons.

The total energy absorbed during the ionization process $Nh\nu$
needs to be divided between ions (as recoil energy), electrons (drift energy) and light (Ponderomotive energy)~\cite{Smeenk11b}.
Considering that the recoil energy to the ion has no polarization dependence, and the pondermotive energy is the same
for linear and circularly polarized light of the same energy, the difference between the number of photons absorbed
depends on the electron drift velocity, which depends strongly on the laser field.
To the drift velocity acquired by the electron correspond
a kinetic energy that has to be supplied by the laser field.
The drift velocity of the electrons (charge $e$, mass $m$) ionized with circularly polarized light (field amplitude $E_0$
and angular frequency $\omega$) is centered at
$e E_0/(m \omega)$~\cite{Zhou11,Smeenk11b}, with a distribution
that corresponds to quantum mechanical tunneling width~\cite{Arissian10}.
For the intensity of $2.8 \times 10^{14}$ W/cm$^2$ that is launched from vacuum into
atmosphere, this electron kinetic energy (due to the drift velocity)
corresponds to 10 photons of infrared. Note that almost the same number of infrared photons
are absorbed to overcome the ionization potential of nitrogen.
For the clamped intensity in filaments however, this kinetic energy (in circular polarization) represents
at most one or two photons.
In linearly polarized light and laser intensities of our experiment, the drift velocity of electrons is centered around zero.
One expect therefore, in circular polarization (as compared to linear polarization), that
a sudden high attenuation of the beam
entering the atmosphere could prevent filamentation.
This effect is augmented by the fact that, at these
intensities, the ionization probability in circular polarization is comparable to the linear one~\cite{Ammosov97}.
In the case of pre-filamentation in air, the ``clamping'' intensity is never exceeded,
and the electron kinetic energy is negligible.

Our observation suggests that the definition of electron index of refraction needs to be revisited.
The generated electron plasma through tunnel ionization can neither be treated as bound or free electrons. At the filament intensities
the tunneled electrons are free from their parent and their excursion is set by the laser field~\cite{Corkum93}.
Their motion is also affected by the Coulomb field of the parent ion,
as they can not move far from the ion in a sub picosecond time scale.
The electron index of refraction is therefore determined by the radiation of
a moving charge~\cite{Jackson75} (Chapter 14), rather than through a steady state approach
that assigns a time independent permittivity $\epsilon $ as a characteristic of the medium.

We note that this approach is very different from the Drude model, which assumes
a uniform isotropic plasma. The number of these electrons also changes rapidly over the pulse.
With this approach the modification of light through electron plasma
needs to be considered at the time scale of the laser pulse.
This applies to the single pulse filament, which is the case of our experiment.
For pump-probe filaments the dynamic initiated by the pump pulse needs to be considered over the delay time.

Since we would like to understand the mechanism of generating a single filament,
we only consider the electron motion during the laser pulse.
The accelerating electrons radiate, a radiation that adds to the applied field.
The low frequency component of this radiation corresponds to acceleration due to the drift velocity,
giving rise to THz radiation.
A 180$^{\rm o}$ out of phase component of the electron re-radiation at the optical light frequency
represents the absorption, or energy lost by the applied field to accelerate the electrons.
A 90$^{\rm o}$ out of phase (advance) component of the electron re-radiation at the optical light frequency
is a contribution of the electron to the index of refraction.
The coherent radiation of electrons added to the applied light field~\cite{Jackson75}(Chapter 14)
provides a full formulation of electron index of refraction.

The difference between filament preparation in air and vacuum not only lies in the intensity provided at the starting point
of filamentation but also on the status of the light polarization.
With pre-filamentation in vacuum we are assured that no mechanism has modified the light spatial profile,
spectrum \cite{Arissian12c} and polarization
prior to reaching the focus.
In traditional filament studies there is no confirmation that the
light in the filament has preserved its initial polarization.
It is a well known fact that, when molecular re-orientation dominates the Kerr effect
(which is the case with 200 fs pulses),
circular polarization is unstable and becomes linear with propagation~\cite{Close66}.
Because of the high intensities inside filaments, it is not easy to determine
their polarization without damaging expensive optics.
It is possible that the 200 fs circularly polarized pulses became
linearly polarized when reaching filament state.
Experiments are in progress to determine
the modification of the polarization due to molecular alignment in case of circularly polarized light
in laser filaments. It should be noted that this alignment has an impact on the
plasma electron density.

In conclusion, we present an experimental investigation of polarization dependence
of filamentation.
Our results add new physics to existing light propagation equations for filamentation~\cite{Couairon07}
which do not include the microscopic effects of electron index and therefore don't show any dependence on light polarization.
We show that the common representation of a uniform electron bubble is an oversimplification
that does not take into account the anisotropy of the electron plasma created by tunnel ionization,
nor the re-radiation of electrons accelerated along different trajectories by linear or
circular polarization, nor the impact of the large drift velocity imparted
to electrons in circular polarization.

Our experimental configuration
eliminates nonlinear propagation effects that precedes filamentation through
pre-focusing in vacuum, and launching the focused beam into atmosphere through an aerodynamic window.
This is in contrast with the situation in various reports on polarization dependence in filament formation such as less supercontinum
generation with circularly polarized light~\cite{Sandhu00, Yang05} and more energy and stability
in circularly polarized filaments in argon\cite{Trisorio07}.

Our results call for a fundamental investigation of the
details of the mutual interaction of radiation and moving charges,
for various light wavelengths and pulse durations, for a well defined light polarization at the focus.
We are indebted to Dr. Paul Corkum and Chris Smeenk of NRC/Ottawa for fruitful discussions.
This work was supported by DTRA under grant number HDTRA1-11-1-0043, and by
the Army Research Office, under the MURI grant
W911NF-11-0297.



\bibliography{c:/bib/ad,c:/bib/en,c:/bib/oz,c:/bib/refbook}
\end{document}